\documentstyle[12pt,epsfig]{article}

\begin{document}

{

\title{Deuteron disintegration by protons
in kinematics of the quasi-elastic
backward $pd$-scattering}
\author{A.~V.~Smirnov\thanks{P.~N.~Lebedev Physical Institute RAS}
 \and Yu.~N.~Uzikov}
\date{}
\maketitle

\centerline{\it Joint Institute for Nuclear Research}

\vspace{0.3cm}
\centerline{\bf Abstract}

The deuteron break-up exclusive reaction
$p+d\to N(180^{\circ})+(NN)(0^{\circ})$ is considered
in the framework of the one-nucleon-exchange mechanism
at the incident proton kinetic energy 0.2--2~GeV and the relative
energy of nucleons in the forward going final $NN$-pair $E_{NN}=0$--20~MeV,
that is close to the kinematical conditions of the planned experiment at COSY.
The off-energy-shell effects resulting, in particular, in the
node of the half-off-shell $^1S_0$-state partial $NN$-scattering
amplitude and the role of higher partial waves
in the subprocess $pN\to pN$ are analysed and found to be very
important, especially for the reaction $p+d\to
n(180^{\circ})+(pp)(0^{\circ})$. The accuracy of some approximate
relations between the cross sections of elastic $pd$-scattering and
deuteron break-up is examined in the framework of the  mechanism
in question.

\eject
\section{Introduction}
The main goal of experiments on deuteron break-up by electrons~\cite{bosted}
or protons~\cite{ableev} at high transferred momenta
consists in getting information
about nucleon interaction at small distances or, in other words,
high-momentum components of the $NN$-wave functions (w.f's).
From this point of view the pole one-nucleon-exchange (ONE) mechanism (Fig.~1)
seems mostly interesting, since in this case the break-up cross
section is proportional to the product of the modules squared of
the deuteron w.f., $|\Psi_d({\bf q})|^2$, where ${\bf q}$ is the relative
$pn$-momentum, and the half-off-shell
$NN$-scattering amplitude, $|T_{NN}({\bf q}',{\bf k})|^2$
( ${\bf q}'$ and ${\bf k}$ are the corresponding relative
momenta in the initial and final $NN$-states, respectively).
For exact definition of $\bf q$, ${\bf q}'$ and $\bf k$ see
Eqs.~(\ref{qd})--(\ref{p}) below.

In previous theoretical approaches to the reactions
$p+d\to p+n+p $~\cite{perdrisat}
and $d+p\to p(0^{\circ})+X$ (see, for instance,~\cite{lykasov},~\cite{ksg}
and references therein) in the framework of the same pole mechanism
the on-shell amplitude of the $pn$-interaction has been used
for the upper vertex, that corresponds to the so-called impulse approximation.
At small values of nucleon-spectator momenta $p'_3<50$~MeV/c
this approximation
looks quite justified, since the departure of the amplitude
$T_{NN}({\bf q}', {\bf k})$ from the energy shell is negligible:
${q'}^2-k^2\approx {p'}_3^2+m|\varepsilon|\ll q'^2,\,k^2$  ($\varepsilon$
is the deuteron binding energy and $m$ denotes the nucleon mass).
 However, when the momentum $p'_3$
is large, validity of the impulse approximation becomes
doubtful and requires special consideration.
\begin{figure}[htb]
\mbox{\epsfig{figure=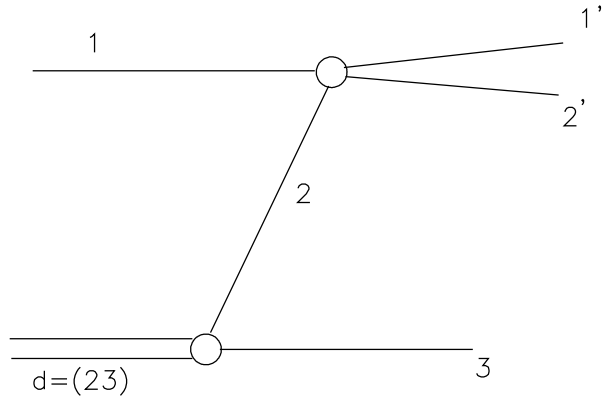,height=0.45\textheight, clip=}}
\caption{One-nucleon-exchange (ONE) mechanism for the reaction
{$p+d\to N+(NN)$}.
}
\end{figure}
The influence of the off-energy-shell effects on the observables for
the reaction of deuteron disintegration by protons
$p+d\to N+(NN)$
in the region of incident proton kinetic energies in the laboratory
frame $T_p=0.5$--2~GeV
was investigated for the first time in Ref.~\cite{imuz90}.
Kinematical conditions of this reaction were chosen
in Ref.~\cite{imuz90} to be similar to those for
quasielastic backward $pd$-scattering, namely, the final nucleon pair
$(NN)$ moved
forward with small relative energy of nucleons, $E_{NN}\leq 3$~MeV,
while the nucleon-spectator
flew in the back half-sphere with almost the same momentum, as for
the initial proton in the $p+d$ center-of-mass frame (CMF). In
Ref.~\cite{imuz90} it was found, that if the  final $NN$-pair is
in the singlet $^1S_0$ state (this takes place for the
reactions $p+d\to n+(pp)$ or $n+d\to p+(nn)$ at $E_{pp}$ and $E_{nn}$
of the order of a few MeV), the contribution  of $\Delta$-isobar mechanism
is considerably suppressed by isotopic factors, whereas
the ONE-mechanism dominates. It was established also, that
the off-energy-shell effects were of great importance in the kinematics
considered. In particular, the differential
cross section of this reaction, calculated in the
framework of the ONE mechanism,
must possess by distinct minimum at
the incident proton energy $T_p\sim 0.7$~GeV.
The occurence of such a minimum is tightly connected with
the node of the half-off-shell $^1S_0$-partial amplitude of $NN$-scattering,
$t(q',k)$ ($q'^2/m\neq k^2/m=E_{NN}$), at the point $q'\approx 0.4$~GeV/c.
This node, in its turn, arises due to repulsive core in the
$^1S_0$-potential of $NN$-interaction, like the node in the
$S$-wave component of the deuteron w.f. in the momentum space
is caused by analogous core in the  $^3S_1$--${^3D_1}$ state.
So, we may conclude that the upper vertex
of the pole diagram of Fig.~1 is not a bit less interesting, than
its lower $d\to p+n$ vertex, since both of them directly relate to
the properties of the $NN$-interaction potential at short
interparticle distances.

In this paper the approach of Ref.~\cite{imuz90} is extended
for including higher partial waves in the amplitude of
$pN$-scattering within the ONE mechanism. This allows to
study the dependence of the deuteron break-up cross section
on the final $NN$-pair relative energy in the range $E_{NN}=0$--20~MeV,
corresponding to conditions of the planned experiment at COSY~\cite{cosy}.
Besides that, we investigate the relations between differential
cross sections of the elastic backward $pd$-scattering and triplet
$pn$-pair formation $p+d\to p+(pn)_t$ on the basis of theorems
 concerning analytic continuation of $pn$-scattering amplitude~\cite{adm}
 and $pn$-scattering w.f.~\cite{bfw} to the bound state pole.

\section{Differential cross section
of deuteron disintegration
in the framework of the ONE mechanism}

We start from the general formula for the invariant
cross section of a process
\begin{equation}
\label {smu1}
p({\bf p}_1,\sigma_1)+d({\bf
P}_d,\lambda_d)\to N({\bf
p}'_1,\sigma'_1) + N({\bf p}'_2,\sigma'_2)
+N({\bf p'}_3,\sigma'_3)
\end{equation}
(in round brackets the
three-momenta and polarizations of the particles are indicated):
\begin{equation}
\label{breakupcs}
d\sigma =(2\pi )^4\delta^{(4)}(P_i-P_f){1\over 4I_0}
{\overline {|A_{fi}|^2}}{d^3p_1'\over {2E_1'(2\pi)^3}}
{d^3p_2'\over {2E_2'(2\pi)^3}}{d^3p'_3\over {2E'_3(2\pi)^3}},
\end{equation}
where
$\overline{|A_{fi}|^2}$ is the spin-averaged amplitude squared of
the reaction given,
$P_i$, $P_f$ are the total four-momenta of the initial and final
systems of particles, respectively,
$I_0=\sqrt{(E_1E_d-{\bf p}_1{\bf P}_d)^2-m^2\, M^2_d}$,
$M_d$ is the deuteron mass,
$E_i=\sqrt{{\bf p}_i^2\mathstrut+m^2}$,
$E_i'=\sqrt{{\bf p}_i^{'2}\mathstrut+m^2}$,
 $E_d=\sqrt{{\bf P}_d^2\mathstrut+M^2_d}$. Performing the integration
over the three-momentum ${\bf p}'_2$ and over the energy $E'_3$, we
get:
\begin{equation}
\label{su3}
E'_1{d\sigma \over d^3p'_1\, d\Omega_3'}=
{1\over {(4\pi )^5}} \,
\frac{{p'}^3_3\,\overline {|A_{fi}|^2}}
{I_0\,\left|E'_2{p'}_3^2-({\bf p}'_2{\bf p}'_3)E'_3\right |}.
\end{equation}

The amplitude $A_{fi}$ is calculated on the basis of concrete
assumptions about dynamics of the process.
ONE is the simplest possible mechanism of the deuteron break-up
reaction. However, as has been mentioned above, it contains
a nontrivial physical information about $NN$-interaction
and, therefore, must be taken into account as a first
approximation.

A correct calculation of the amplitude of the ONE diagram (Fig.~1) requires
incorporation of relativistic effects.
Unfortunately, at present the problem of describing reativistic processes
with hadronic  compound systems is not completely solved yet.
Different approaches give different
approximate solutions coinciding only in the nonrelativistic limit.
For instance, a widely spread method based on the noncovariant
light-front dynamics or equivalent to it dynamics in the infinite
momentum frame (see Ref.~\cite{lykasov} and references therein) suffers
from explicit violation of the rotational invariance, that becomes
a definite flaw at intermediate energies. A more general approach ---
the covariant light-front dynamics~\cite{karm} --- turns out
to be very effective
for solving the problems of electromagnetic interactions with
relativistic bound systems, including deuteron~\cite{karmsm}.
However, its practical
application to the processes of hadron-deuteron interactions still
encounters some difficulties connected with dependence
of approximately calculated amplitudes on the position of the light
front surface. Using the covariant Feynman technique requires extremely
complicated calculations
of the deuteron Bethe--Salpeter function and the half-off-mass-shell
$NN$-amplitude. Although such an algorithm has been partially realized
in Ref.~\cite{kaptari} for the inclusive reaction $p+d\to p(0^{\circ})+X$,
the simplifying approximations made in that paper (neglect of the spin
structure and off-shell effects in the $NN$-scattering amplitude) are
evidently not applicable for our purposes.

Keeping all these in mind, we apply here for constructing
the amplitudes $p+d\to N+N+N$ and $p+d\to p+d$ the same method, as
was used in Ref.~\cite{imuz90}, namely, the relativistic quantum
mechanics of three-body system~\cite{bkt}, based on the construction of
the full set of the P\'{o}incare group generators.
In this approach the equation for two-body mass operator
eigenfunctions and eigenvalues coincides in form with the
Schr\"{o}dinger equation~\cite{coester}. In spite of restriction by the sector
with fixed number of particles (that is, in fact, a minimal
relativisation only), such an approach allows to incorporate in
a self-consistent way a rather rich nonrelativistic phenomenology of
$NN$-interactions.

In the framework of the formalism~\cite{bkt} the expression for
the ONE-amplitude
of the process~(\ref{smu1}) has the simplest form in the CMF of the system
$p+d$~\cite{uz92}
\begin{equation}
\label{breakupamp}
A_{fi}=
 {\sqrt{E_d(E_2+E_3')\, \varepsilon_p(q)}\over E_2}
\sum_{\sigma_2}
\Psi^{\lambda_d}_{\sigma'_3 \, \sigma_2}({\bf q})
T_{NN\,\sigma _1\,\sigma_2}^
{\hphantom{NN}\sigma' _1\,\sigma'_2}({\bf q}', {\bf k}).
\end{equation}
Here $\varepsilon_p(q)=\sqrt{{\bf q}^2+m^2\mathstrut}$ and
the relativistic relative momenta ${\bf q}$, ${\bf q}'$ and ${\bf k}$
are defined as
\begin{equation}
\label{qd}
{\bf q}=L^{-1}\left( \frac{m({\bf p}_2+{\bf p}'_3)}
{\sqrt{(p_2+p'_3)_{\mu}^2}}\right){\bf p}'_3,
\end{equation}
\begin{equation}
\label{k}
{\bf q}'=L^{-1}\left(\frac{m({\bf p_1}+{\bf p}_2)}
{\sqrt{(p_1+p_2)_{\mu}^2}}\right){\bf p}_1,
\end{equation}
\begin{equation}
\label{p}
{\bf k}=L^{-1}\left(\frac{m({\bf p}'_1+{\bf p}'_2)}
{\sqrt{(p'_1+p'_2)_{\mu}^2}}\right){\bf p}'_1,
\end{equation}
where ${\bf p}_1,\, {\bf P}_d, \,{\bf p}'_1, \,{\bf p}'_2, \,
{\bf p}'_3 \, $ are taken in the CMF of $p+d$, where, according
to Ref.~\cite{bkt}, the conservation of 3-momenta in the vertices
$d\to 2+3'$
and $1+2\to 1'+2'$ takes place. By this reason
we use in  Eqs.~(\ref{qd}),~(\ref{k}) the relations
${\bf p}_2={\bf P}_d-{\bf p}'_3$ and $(p_2)_0\equiv E_2=\sqrt{{\bf p}_2^2
\mathstrut +m^2}$. $(p_i)_{\mu}$ is the four-momentum of the particle $i$
and $L^{-1}({\bf Q})$ is the Lorentz boost acting on some
four-momentum $a_{\mu}=(a_0,{\bf a})$, $a_{\mu}^2=m^2$ as follows:
\begin{equation}
\label{lorentz}
L^{-1}({\bf Q}){\bf a}={\bf a}-{{\bf Q}\over m}\,\left[
a_0-\frac{{\bf a}{\bf Q}}{m+\sqrt{{\bf Q}^2\mathstrut +m^2}}\right].
\end{equation}
By its physical meaning, the relative momentum
of a pair of nucleons is nothing else, than
the momentum of one of the nucleons in their common CMF.
The momenta~(\ref{qd}) and~(\ref{k}) are completely identical to
the corresponding momenta ${\bf q}$ and ${\bf q}'$
defined by the formulas~(17),~(18) in Ref.~\cite{imuz90}.
In the nonrelativistic limit we get ${\bf q}=({\bf p}'_3-{\bf p}_2)/2$,
${\bf q}'=({\bf p}_1-{\bf p}_2)/2$, ${\bf k}=({\bf p}'_1-{\bf p}'_2)/2$,
and the kinematical factor
${\sqrt{E_d(E_2+E_3')\, \varepsilon_p(q)}/ E_2}$ in
Eq.~(\ref{breakupamp}) turns to $2\sqrt{m}$, as it ought to be.

The half-off-shell $NN$-scattering amplitude
$T_{NN\,\sigma _1\,\sigma_2}^
{\hphantom{NN}\sigma'_1\,\sigma'_2}({\bf q}', {\bf k})$
in Eq.~(\ref{breakupamp}) satisfies
the Lippman--Schwinger equation
\footnote{According to Ref.~\cite{bkt} the amplitude $T_{NN}$
coming into the formula~(\ref{breakupamp})
is the two-body amplitude of the process $1+2\to 1'+2'$
in the three-body space. In contrast to the nonrelativistic
approach, it does not coincide with the two-body amplitude for
 the purely two-body problem.
The coincidence can be reached, strictly speaking, only in the limit
$(p_1+P_d)^2_{\mu}\to \infty$ (see Ref.~\cite{uz92}). We will neglect this
circumstance, considering the amplitude $T_{NN}$ as obeying the two-body
Lippman--Schwinger equation.}
$$
T_{NN\,\sigma _1\,\sigma_2}^
{\hphantom{NN}\sigma' _1\,\sigma'_2}({\bf q}', {\bf k})=
-4m^2U^{\sigma'_1\,\sigma'_2}_{\sigma_1\,\sigma_2}({\bf q}'-{\bf k})$$
\begin{equation}
\label{ls}
-m\sum_{\rho_1\,\rho_2}\int\frac{d^3p}{(2\pi)^3}\,
\frac{U^{\sigma'_1\,\sigma'_2}_{\rho_1\,\rho_2}({\bf q}'-{\bf p})
T_{NN\,\sigma _1\,\sigma_2}^
{\hphantom{NN}\rho_1\,\rho_2}({\bf p}, {\bf k})}
{p^2-k^2-i0},
\end{equation}
where $U$ is the Fourier transform of the interaction potential.
Being taken on the energy shell $q'=k$, the amplitude $T_{NN}$
relates to the CMF differential cross section of unpolarized nucleons as
\begin{eqnarray}
\label{crosssection}
\frac{d\sigma_{\rm 12}}{d\Omega}=\frac{1}{64\pi^2 s_{\rm 12}}\,\frac{1}{4}\,
\sum_{\sigma'_1\,\sigma'_2\atop \sigma_1\,\sigma_2}
\left|T_{NN\,\sigma _1\,\sigma_2}^
{\hphantom{NN}\sigma' _1\,\sigma'_2}({\bf q}', {\bf k})
\right|^2.
\end{eqnarray}
Here $s_{\rm 12}=4({\bf q}'\,^2+m^2)=4({\bf k}^2+m^2)$. The factor 1/4 in
Eq.~(\ref{crosssection}) appeared because of averaging over initial
particle polarizations.

The w.f. $\Psi^{\lambda_d}_{\sigma_3'\,\sigma_2}({\bf q})$ is
the eigenfunction of the mass squared operator of two interacting nucleons,
corresponding to the eigenvalue $M_d^2$.
It satisfies the following equation~\cite{coester}:
\begin{equation}
\label{sh}
\left(\frac{q^2}{m}+ m-\frac{M_d^2}{4m}\right)
\Psi^{\lambda_d}_{\sigma_3'\,\sigma_2}
({\bf q})+\sum_{\rho_1\,\rho_2}\int \frac{d^3p}{(2\pi)^3}\,
U^{\sigma_3'\,\sigma_2}_{\rho_1\,\rho_2}({\bf q}-{\bf p})
\Psi^{\lambda_d}_{\rho_1\,\rho_2}({\bf p})=0
\end{equation}
and the normalization condition
\begin{equation}
\label{norm}
\sum_{\sigma_2\,\sigma_3'}
\int {d^3q\over (2\pi)^3}\,\Psi^{\lambda'_d\,*}_{\sigma_3'\,\sigma_2}
({\bf q}) \Psi^{\lambda_d\hphantom{\lambda'_d*}}
_{\sigma_3'\,\sigma_2}({\bf q})=\delta_{\lambda'_d\,\lambda_d}.
\end{equation}
The quantity $(M_d^2/4m)-m$ in Eq.~(\ref{sh}) is very close to the
deuteron binding energy $\varepsilon=M_d-2m$ and,
consequently, the w.f.  $\Psi^{\lambda_d}_{\sigma_3'\,\sigma_2}({\bf
q})$ practically coincides with the solution of the Schr\"{o}dinger
equation in the momentum space.

For further analysis it is convenient to introduce partial $NN$-scattering
amplitudes $t^{JS}_{LL'}(q',k)$ by means of the following decomposition:
$$T_{NN\,\sigma _1\,\sigma_2}^
{\hphantom{NN}\sigma' _1\,\sigma'_2}({\bf q}', {\bf k})=
4\pi\sum_{JM_JLL'S\atop m_Lm_L'm_Sm_S'}
{\cal N}_{NN}(L,S)\, C^{JM_J}_{Lm_L\,Sm_S}\,
C^{JM_J}_{L'm_L'\,Sm_S'}$$
\begin{equation}
\label{t}
\times C^{Sm_S}_{{1\over 2}\sigma_1\,{1\over 2}\sigma_2}\,
C^{Sm_S'}_{{1\over 2}\sigma'_1\,{1\over 2}\sigma'_2}\,
Y_{Lm_L}^*(\hat{{\bf q}'}\,)Y_{L'm_L'}(\hat{{\bf k}}\,)
\,t^{JS}_{LL'}(q',k),
\end{equation}
where $C^{j_1m_1}_{j_2m_2\,j_3m_3}$ are Clebsch--Gordan coefficients
and
the combinatorial factor
\begin{equation}
\label{comb}
{\cal N}_{NN}(L,S)=\left\{
\begin{array}{ll}
1+(-1)^{L+S}, & \mbox{if}\,\,\,NN=pp\,\,\mbox{or}\,\,nn,\\
1, & \mbox{if}\,\,\,NN=pn\\
\end{array}
\right.
\end{equation}
is introduced to take into account antisymmetric properties of the
amplitude
for the case of identical nucleons.
Indeed, for $pp$- or $nn$-scattering the states with even $L+S$ only
are admittable, that follows from the Pauli principle, while for
nonidentical nucleons the sum $L+S$ can be arbitrary, and no
restrictions
on the quantum numbers are imposed.

We represent the deuteron w.f. analogously to Eq.~(\ref{t}):
\begin{equation} \label{sd}
\Psi_{\sigma_3'\,\sigma_2}^{\lambda_d}({\bf q})= \sum_{L=0,2\atop
\mu_d\,m_L} C^{1\lambda_d}_{Lm_L\,1\mu_d}\,C^{1\mu_d}_{{1\over
2}\sigma_3'\, {1\over 2}\sigma_2}\,Y_{Lm_L}(\hat{{\bf q}})\,u_L(q).
\end{equation}
Here $u_0(q)$ and $u_2(q)$ are the usual $S$- and $D$-waves in deuteron,
normalized as
\begin{equation}
\label{normsd}
\int _0^\infty {dq\,q^2\over (2\pi)^3}\,\left[u_0^2(q)+u_2^2(q)\right]=1.
\end{equation}
Substituting Eqs.~(\ref{t}),~(\ref{sd}) into Eq.~(\ref{breakupamp}),
we get after squaring, averaging over initial and summation over final
polarizations
\begin{eqnarray}
\label{amplsquared}
{\overline {|A_{fi}|^2}}= {1\over 6}\sum_ {\lambda_d\,\sigma_1
\, \sigma_1' \,\sigma_2'\,\sigma'_3} {|A_{fi}|^2} \nonumber \\
={E_d(E_2+E_3')\, \varepsilon_p(q)\over 16\pi E_2^2}
\left[u_0^2(q)+u_2^2(q)\right]\,
 F({\bf q}',{\bf k}).
\end{eqnarray}
The function $F({\bf q}',{\bf k})$ can be written in the following form:
$$F({\bf q}',{\bf k})=
\sum_{\sigma'_1\,\sigma'_2\atop \sigma_1\,\sigma_2}
\left|T_{NN\,\sigma _1\,\sigma_2}^
{\hphantom{NN}\sigma' _1\,\sigma'_2}({\bf q}', {\bf k})
\right|^2$$
$$
=\sum_{JLL'\widetilde{J}\widetilde{L}\widetilde{L'}Sl}
 {\cal N}_{NN}^2(L,S)\, t_{LL'}^{JS}(q',k)
\left (t_{\widetilde {L}\widetilde {L'}}^{\widetilde {J}S}(q',k)\right )^*
\,C^{\widetilde{L}0}_{L0\,l0}\,C^{\widetilde{L'}0}_{L'0\,l0}
(2J+1)(2\widetilde{J}+1)$$
\begin{equation}
\label{fpq}
\times (2l+1)\sqrt{(2L+1)(2L'+1)}
\left \{ \begin{array} {ccc}{\widetilde L}&L&l\\ J&{\widetilde J}&S
\end{array}\right \}
\left \{ \begin{array} {ccc}{\widetilde L}'&L'&l\\
 J&{\widetilde J}&S \end{array}\right \}\,P_l({\bf q}'{\bf k}/q'k),
\end{equation}
where the figured brackets stand for usual notation of $6j$-symbols.
Note that, according to Eq.~(\ref{crosssection}), the
function~(\ref{fpq}) at $q'= k$ describes the differential cross
sections of $pn$- and $pp$-scattering.
As is seen from Eq.~(\ref{fpq}), the states with different spins $S$
do not interfere in the cross section of the reaction, that is
a consequence of the total
angular momentum $J$ and parity conservation.  Therefore, it makes
sense to separate transitions in singlet ($S=0$) and triplet ($S=1$)
channels.

The partial amplitudes $t^{JS}_{LL'}(q',k)$ are found by means
of numerical solving the
Lippman--Schwinger equation~(\ref{ls}), which after separation of
angular variables reduces to the system of coupled one-dimensional
integral equations \begin{equation} \label{lsrad}
t^{JS}_{LL'}(q',k)=v^{JS}_{LL'}(q',k)+\frac{2}{(4\pi)^2}\,
\sum_{L''}\int^{\infty}_0\frac{dp\,p^2}{m}\,\frac{v^{JS}_{LL''}(q',p)
t^{JS}_{L''L'}(p,k)}{p^2-k^2-i0},
\end{equation}
where
$$v^{JS}_{LL'}(q',k)=-\frac{m^2}{\pi}
\sum_{\sigma_1,\sigma_2,\sigma'_1,\sigma'_2\atop m_S,m'_S,m_L,m_L'}
\int d\Omega_{{\bf q}'}d\Omega_{\bf k}\,
C^{JM_J}_{Lm_L\,Sm_S}\,
C^{JM_J}_{L'm_L'\,Sm_S'}\,
C^{Sm_S}_{{1\over 2}\sigma_1\,{1\over 2}\sigma_2}\,
C^{Sm_S'}_{{1\over 2}\sigma'_1\,{1\over 2}\sigma'_2}$$
\begin{equation}
\label{potential}
\times U^{\sigma'_1\,\sigma'_2}_{\sigma_1\,\sigma_2}({\bf q}'-{\bf k})
\,Y_{Lm_L}(\hat{{\bf q}}')\,Y_{L'm_L'}^*(\hat{{\bf k}}).
\end{equation}

Concerning the reaction $p+d\to n+p+p$ at small relative energy
of the final protons
the question about the role of the Coulomb effects in $pp$-scattering
arises.
To take the Coulomb interaction into account, it is necessary to
replace in Eqs.~(\ref{ls}),~(\ref{potential}) the function
$U$ by the sum $U+U_C$,
where $U_C$ is the Fourier transform of the Coulomb potential.
However, as was found in Ref.~\cite{imuz90}, for a rather big
departure from the energy shell, $q'\gg k$, the influence of the
Coulomb forces becomes negligible. Since this just takes place
for our case, we will exclude the Coulomb effects from consideration.
Let us emphasize, that on the energy shell $q'=k$ such an approximation
would be unacceptable.

If the kinematics is collinear (the scattering angles in the
laboratory frame are $\theta'_1=\theta'_2=0^{\circ}$ for nucleons
in the $NN$-pair, $\theta'_3=180^{\circ}$ for the nucleon-spectator),
then ${\bf q'}\parallel {\bf k}$, and
Eq.~(\ref{fpq}) can be simplified:
\begin{equation}
\label{fsimpl}
F({\bf q}',{\bf k})
=|A_0|^2+\sum_{\mu=-1,0,1}|B_{\mu}|^2
\end{equation}
with
\begin{equation}
\label{ab0b1}
A_0=\sum_{J}{\cal N}_{NN}(J,0)\,\xi_J\,(2J+1)t^{J0}_{JJ}(q',k),
\end{equation}
\begin{equation}
\label{b0b1}
B_{\mu}=\sum_{JLL'}{\cal N}_{NN}(L,1)\,\xi_L\,
(2J+1)C^{L0}_{J\mu\,1\,-\mu}
C^{L'0}_{J\mu\,1\,-\mu}
\,t^{J1}_{LL'}(q',k)
\end{equation}
and
\begin{equation}
\label{xi}
\xi_l=\left\{{1,\,\,\,\,\,\,\,\,\,\,\,\mbox{if}\,\,\,({\bf q}'{\bf k}/q'k)=1,
\atop
(-1)^l,\,\,\,\,\mbox{if}\,\,\,({\bf q}'{\bf k}/q'k)=-1.}\right.
\end{equation}
The terms $|A_0|^2$ and $|B_{\mu}|^2$ in Eq.~(\ref{fsimpl}) represent
the contributions from singlet ($S=0$) and triplet ($S=1$, $m_S=\mu$)
states of the $NN$-pair, respectively. Note that $B_{-1}=B_1$.

Provided the relative energy of nucleons in the pair is small enough,
we have a quasi two-body  kinematics.
However, the state of this pair is characterized by spin projections
$\sigma'_1$ and $\sigma'_2$ of individual nucleons, rather than by definite
common quantum numbers like the total angular momentum $J$, its projection
$M_J$,
total spin $S$ and orbital momentum $L$ (as it will be
discussed in the next section.)
\begin{figure}[ht]
\mbox{\epsfig{figure=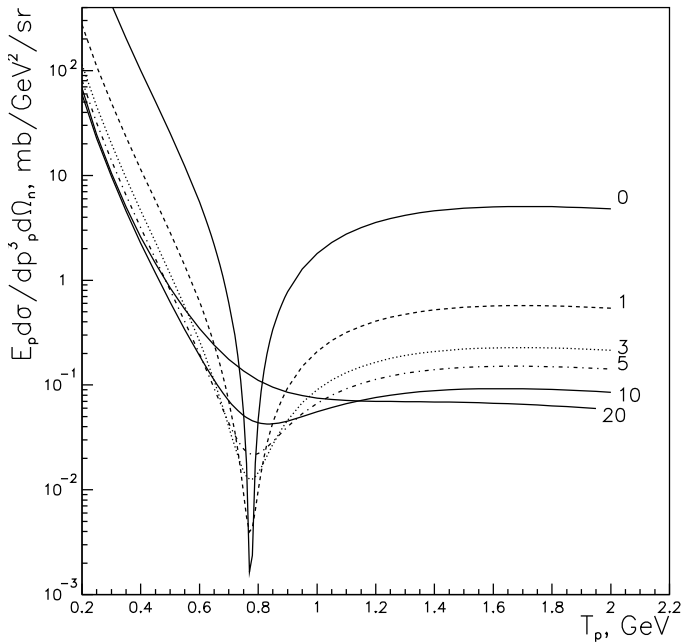,height=0.45\textheight, clip=}}
\caption{ The CMF differential cross section of the reaction
{$p+d\to p(0^{\circ})+p+n(180^{\circ})$} as a function of the
initial proton kinetic energy in
the laboratory frame, calculated in the framework of the ONE mechanism
by Eqs.~(\protect\ref{su3}),~(\protect
\ref{amplsquared}),~(\protect\ref{fsimpl}),~(\protect\ref{ab0b1})
with $S$-, $P$-, $D$-, $F$- and $G$-waves in the state of the
forward going $pp$-pair taken into account.
The numbers at the curves denote the energy of relative motion of
nucleons in the pair, $E_{pp}$, in MeV.}
\end{figure}
Then, below in this section we will denote the reaction considered
as $p+d\to N'_1(0^{\circ})+N'_2+N'_3(180^{\circ})$ to indicate
explicitly, which nucleons are registered in the experiment
(in brackets the corresponding scattering angles of registered
nucleons in the laboratory frame are pointed out). At the same time,
one should remember that the kinematics corresponds to the situation,
when the nonregistered nucleon $N'_2$ moves at the angle $0^{\circ}$
in the laboratory system, and the relative energy of the forward
going pair of nucleons $N'_1N'_2$ is small: $E_{NN}=0$--20~MeV.
As has been noticed above,
in the collinear kinematics discussed the angle between the relative
momenta ${\bf q}'$ and ${\bf k}$
can be either $0^{\circ}$, or $180^{\circ}$.
For the case of $pp$-pair formation the result for the cross section
does not depend on which sign of the scalar product $({\bf q}'{\bf k})$
in Eq.~(\ref{xi}) we take. Indeed, since $\xi_J=1$ for $S=0$ and $\xi_L=-1$
for $S=1$, under the substitution $({\bf q}'{\bf k})\to
-({\bf q}'{\bf k})$ one gets $A_0\to A_0$, $B_{\mu}\to -B_{\mu}$,
and the function $F({\bf q}',{\bf k})$ remains unchanged.
But the function $F({\bf q}',{\bf k})$ for $pn$-pair does not
possess by this property. In principle,
the cases $({\bf q}'{\bf k}/q'k)=-1$ and
$({\bf q}'{\bf k}/q'k)=1$ can be experimentally separated.
Our calculations showed that the replacement of
$({\bf q}'{\bf k}/q'k)=-1$ to
$({\bf q}'{\bf k}/q'k)=1$ at such low relative energies
of $pn$-pair as 0--20~MeV  results in some quantitative
difference in the cross sections obtained, but, nevertheless,
does not lead to appearance of any new qualitative features
in the cross section behaviour. By this reason, below we consider
in detail the case of zero angle between ${\bf q}'$ and ${\bf k}$.

The numerically calculated  differential  cross sections
of the reactions
$p+d\to p(0^{\circ})+p+n(180^{\circ})$ and
$p+d\to p(0^{\circ})+n+p(180^{\circ})$
as functions of the initial proton kinetic
energy in the laboratory frame $T_p$ ($T_p=0.2$--2~GeV) are shown in
Figs.~2,~3, respectively.
The calculations were carried out by the
formulas~(\ref{su3}),~(\ref{amplsquared}),~(\ref{fsimpl})--(\ref{xi}).
For finding the deuteron w.f. and $NN$-scattering amplitudes the Paris
potential~\cite{lacombe} was used. We found that it was enough to
restrict ourselves by taking into account the partial amplitudes
$t^{JS}_{LL'}(q',k)$ describing transitions to the $S$-, $P$-,
$D$-, $F$- and $G$-states of the final nucleon pair (i.~e. those
with $L'=0,\,1,\,2,\,3,\,4$), the relative contribution of the
amplitudes with higher $L'$ being less than 1\% in all the
range of $T_p$ and $E_{NN}$ considered.
\begin{figure}[htb]
\mbox{\epsfig{figure=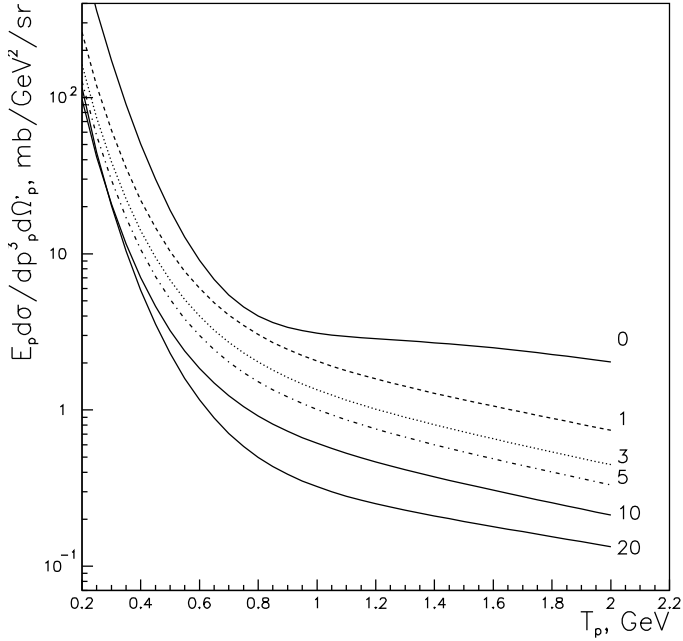,height=0.450\textheight, clip=}}
\caption{ The same as in Fig.~2, but for the reaction
{$p+d\to p(0^{\circ})+n+p(180^{\circ})$}.}
\end{figure}

The first qualitative feature, which draws attention, is a distinct
minimum of
the cross section for $pp$-pair formation
at $T_p=0.77$~GeV (see Fig.~2),
occuring for relative energies
$E_{pp}\sim 0 - 5$~MeV. This minimum is caused by the
node of the half-off-shell amplitude $t^{00}_{00}(q',k)$ corresponding
to the $^1S_0\to {^1S_0}$ transition.
In paper~\cite{imuz90}, where
the Reid-soft-core $NN$-potential was exploited, the
similar minimum took place at another energy $T_p\sim 0.7$~GeV, that
was connected with the difference between the Reid-soft-core and Paris
potentials. At $E_{pp}>5$~MeV the minimum rapidly fills up due
to the contributions from the highest partial waves, and at $E_{pp}>10$~MeV
it practically disappears. For the $pn$-pair formation the behaviour of
the corresponding cross section (Fig.~3)
is, on the contrary, monotonous for all values of $E_{pn}$.
The reason for it consists in the following. At small enough $E_{NN}$
the behaviour of the cross sections discussed
is determined, in the main, by the $NN$-amplitudes $t^{JS}_{L0}(q',k)$
describing transitions to the $S$-states of the final $NN$-pair.
Three such amplitudes exist, namely,
$t^{00}_{00}(q',k)$ $({^1S_0}\to {^1S_0})$,
$t^{11}_{00}(q',k)$ $({^3S_1}\to {^3S_1})$ and
$t^{11}_{20}(q',k)$ $({^3D_1}\to {^3S_1})$.
For the case of $pp$-pair formation
the two latter amplitudes are absent (since for them $L+S$ is odd),
the only remaining $t^{00}_{00}(q',k)$-amplitude has a node leading
to the minimum in the cross section,
as has been already pointed out above.
For the case of $pn$-pair there are no any restrictions on quantum numbers,
and all the three transitions are admittable. The amplitude
$t^{11}_{00}(q',k)$, like $t^{00}_{00}(q',k)$, possesses by a node,
while $t^{11}_{20}(q',k)$ does not. This situation is illustrated
by Fig.~4, where the modules of the three amplitudes considered are
shown as functions of $q'$ at $k=\sqrt{mE_{NN}}=0$. We see that
$|t^{11}_{20}(q',k)|$ has a broad maximum just near the
points of the nodes of $|t^{00}_{00}(q',k)|$ and $|t^{11}_{00}(q',k)|$
(the positions of these nodes are very close to each other).
Namely the contribution of the "nodeless"
amplitude $t^{11}_{20}(q',k)$ leads
to disappearance of the minimum in
the cross section of $pn$-pair production.

Both $pp$- and $pn$-pair formation cross sections
fall sharply
with increase of $T_p$ up to the value $T_p\simeq 0.8$~GeV.
This is caused by the simultaneous decrease of the deuteron
w.f. squared $u^2_0(q)+u_2^2(q)$ and the function $F({\bf q}',{\bf k})$
(Eq.~(\ref{fpq})) in that region.
The nodes of the $t^{00}_{00}(q',k)$- and
$t^{11}_{00}(q',k)$-amplitudes lead to a minimum of $F({\bf q}',{\bf k})$
at $T_p\sim 0.8$~GeV; with further increase of $T_p$ the function
$F({\bf q}',{\bf k})$ starts growing.
For the case of $pp$-pair with
small relative energy, $E_{pp}<10$~MeV, the rate of this growth is
enough to suppress the decrease of the factor $u_0^2(q)+u_2^2(q)$.
As a result, the corresponding cross section in the
region $0.8$~$\mbox{GeV}<T_p<1.4$~GeV increases considerably.
It is interesting to note
that at higher $T_p$ the $pp$-pair formation cross section
becomes practically a constant. For the $pn$-pair the growth of
the function $F({\bf q}',{\bf k})$ is not so fast
and can only reduce the slope of the cross sections.

As is seen from Figs.~2,~3, the absolute values
of the cross sections of both $pp$- and $pn$-pair formation
are very sensitive to
variations of $E_{NN}$, especially at $E_{NN}$ close to zero.
The latter circumstance is stipulated mainly by the virtual state
pole of the $t^{00}_{00}(q',k)$-amplitude at $k^2/m\sim-0.1$~MeV.
On the whole, as $E_{NN}$ increases from 0 to 20~MeV,
both $pp$- (excepting for narrow region near the
point of the minimum) and $pn$-pair formation cross sections reduce
almost by two orders of magnitude.
\begin{figure}[h]
\mbox{\epsfig{figure=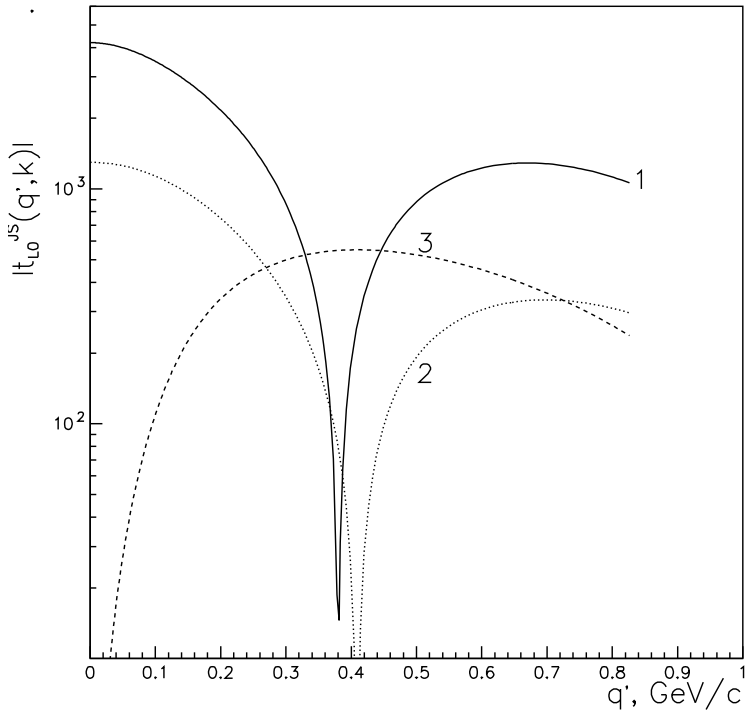,height=0.50\textheight, clip=}}
\caption{
The modules of the half-off-shell partial amplitudes of
$NN$-scattering, $|t^{00}_{00}(q',k)|$ (the solid curve 1),
 $t^{11}_{00}(q',k)$ (the dotted curve, 2),
$|t^{11}_{20}(q',k)|$ (the dashed curve 3)  as functions of $q'$ at $k=0$.}
\end{figure}

At $E_{NN}\to 0$ the partial amplitude $t^{00}_{00}(q',k)$ becomes
dominant (again, unless $q'$ is in the vicinity of the minimum point) due
to its nearthreshold pole mentioned above, which is
much closer to the $NN$-threshold, than the pole of the
$t^{11}_{00}(q',k)$- and $t^{11}_{20}(q',k)$-amplitudes,
occuring at $k^2/m=-2.23$~MeV and corresponding to deuteron.
If we retained on the right-hand-side of Eq.~(\ref{fpq}) the term
with the $t^{00}_{00}(q',k)$-amplitude only, then the ratio of
the $pp$- and $pn$-pair formation cross sections would be exactly
equal to ${\cal N}_{pp}^2(0,0)/{\cal N}_{pn}^2(0,0)=4$
(note that if the final pair
is in the antisymmetric state with quantum numbers $J$, $M_J$, $L$, $S$
then this ratio equals 2~\cite{imuz90}).
In Fig.~5 the cross sections for $pp$- and $pn$-pair formation
(solid lines) are shown together at $E_{pp}=E_{pn}=0$. As is seen,
the cross section for $pp$-pair production is indeed greater
than the one for $pn$-pair anywhere beyond
the region close to the minimum of the former.
However, at $T_p=0.2$--2~GeV their ratio does not exceed 2.77.
This indicates, that the contribution of the $^3S_1$-state
of the $pn$-pair (shown in Fig.~5 by the dotted line)
is not negligible under these conditions even for
$E_{pn}=0$.
Moreover, it is easy to get from
Eqs.~(\ref{su3}),~(\ref{amplsquared}),~(\ref{fsimpl}),~(\ref{ab0b1})
that  at $E_{pn}=0$ the $pn$-pair production cross section is
proportional to the sum
$$|t^{00}_{00}(q',k)|^2+3\left(|t^{11}_{00}(q',k)|^2+
|t^{11}_{20}(q',k)|^2\right),$$
so the contribution of the triplet state of the pair is additionally
amplified by the spin statistical factor of 3.
\begin{figure}[tbh]
\mbox{\epsfig{figure=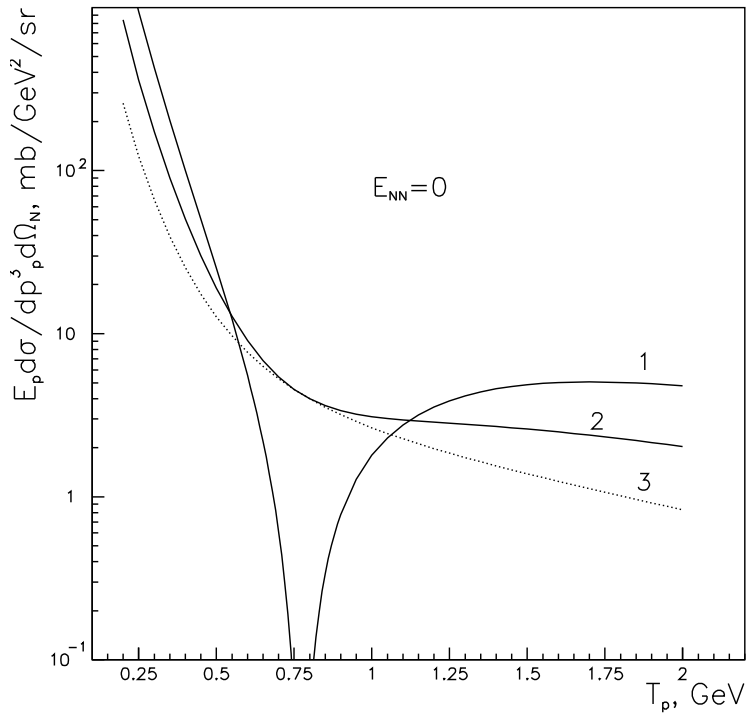,height=0.50\textheight, clip=}}
\caption{
Comparison of contributions of different $S$-states of the
final forward going nucleon pair to the CMF differential cross sections
of the reaction
{$p+d\to p(0^{\circ})+N+N(180^{\circ})$} at $E_{\rm pN}=0$:
the contribution of the $^1S_0$-state for $pp$-pair (the solid curve 1),
total contribution of the $^1S_0$- and $^3S_1$-states for $pn$-pair
(the solid curve 2), the single $^3S_1$-state contribution for $pn$-pair
(the dotted curve 3).}
\end{figure}

To analyse the contribution of the highest partial waves (i.~e. those
with $L'\neq 0$) in the state of the final $NN$-pair, we plotted in
Fig.~6 the cross sections for $pp$- and $pn$-pair formation,
calculated by the same
formulas~(\ref{su3}), (\ref{amplsquared}),~(\ref{fsimpl})--(\ref{xi})
with all the important transition amplitudes $t^{JS}_{LL'}(q',k)$
included (the solid lines~3 and~5)
and with transition amplitudes $t^{JS}_{L0}(q',k)$ only kept
(the dashed lines~2 and~4).
All the curves mentioned correspond to $E_{NN}=20$~MeV.
The influence of the partial waves with $L'\neq 0$ in the $pp$-pair state
is seen to be
very important, especially near the minimum of the dashed curve
for $pp$-pair.
For the case of $pn$-pair the states with $L'\neq 0$ contribute
significantly only at rather low energies $T_p\sim 0.2$--$0.5$~GeV,
increasing the cross section in this region by a factor of $1.25$--$1.5$.
At higher values of $T_p$ the total contribution of the $^1S_0$- and
$^3S_1$-states of the $pn$-pair practically exhausts the cross section
due to destructive interference of the amplitudes with $L'\neq 0$
in Eqs.~(\ref{ab0b1}),~(\ref{b0b1}). For the $pn$-pair case the role of higher
partial waves rapidly reduces
with decrease of $E_{pn}$. For instance,
already at $E_{pn}=10$~MeV the relative error of neglect of the $pn$-pair
states with $L'\neq 0$ is less than 5\% at any $T_p$.
For
$pp$-pair taking into account the highest partial waves near the minimum point
of the $^1S_0$-state contribution
remains important even for much smaller values of $E_{pp}$.
\begin{figure}[hbt]
\mbox{\epsfig{figure=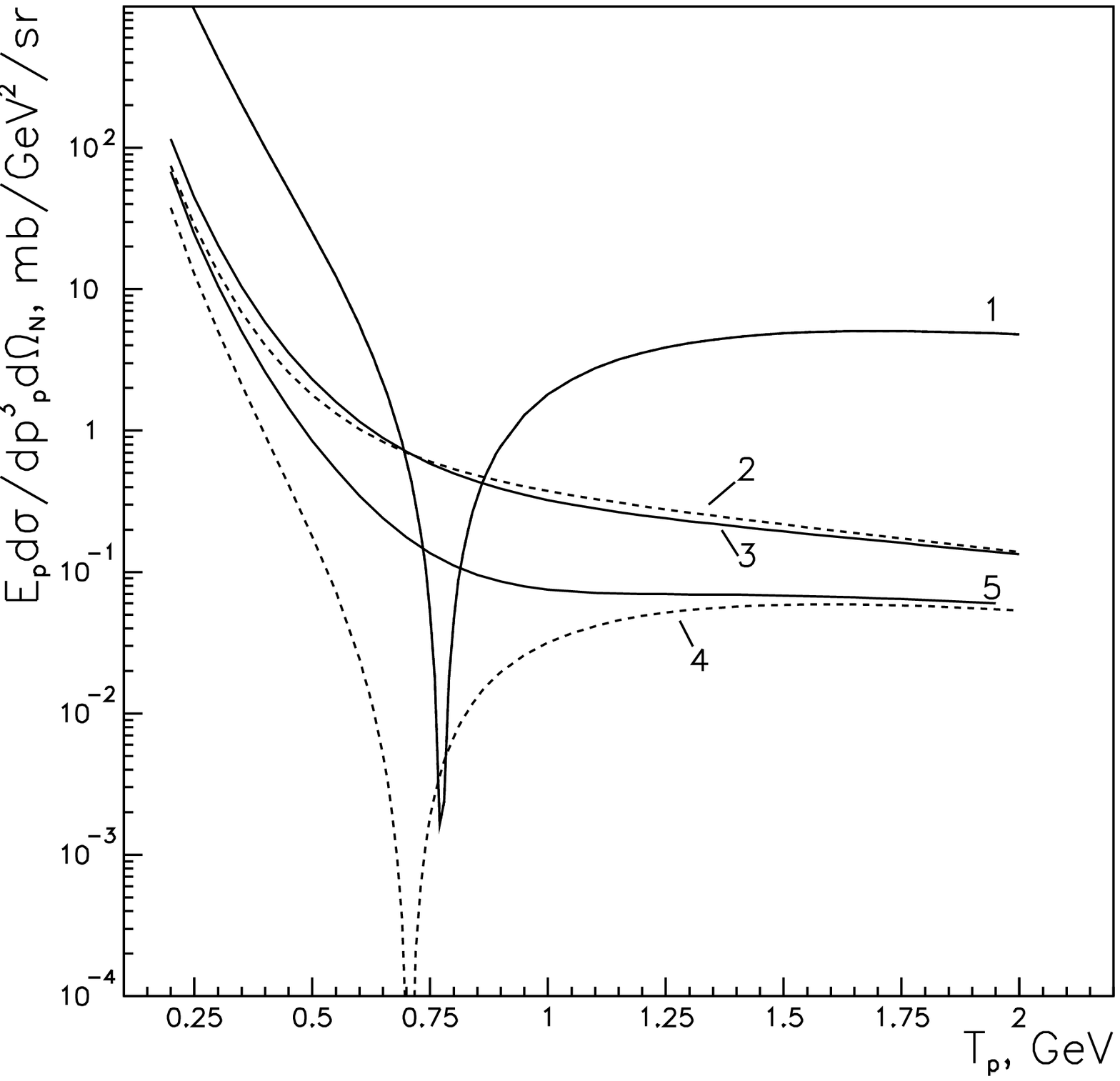,height=0.50\textheight, clip=}}
\caption{ The CMF differential
cross sections of the reaction {$p+d\to p(0^{\circ})+N+N(180^{\circ})$}
for $pp$- and $pn$-forward going pairs formation.
The curves show the results of calculations
taking into account all the important partial waves
(i.~e. $S$-, $P$-, $D$-, $F$-, $G$-) in the state of the pair
(solid lines) and $S$-waves only (dashed lines):
 1 --- for $pp$-pair
at $E_{pp}=0$;
2 and 3 --- for $pn$-pair at $E_{pn}=20$~MeV;
4 and 5 --- for $pp$-pair at $E_{pp}=20$~MeV}


\end{figure}

One more circumstance is also worth to emphasize.
It is known~\cite{plessh} that the position of the node of the amplitude
$t_{00}^{00}(q',k)$ as a function of $q'$ practically does not depend on $k$
in a rather wide interval of $E_{NN}=0$--$100$~MeV.
However, when $E_{NN}$ varies from от 0 to 20~MeV,
the node shifts approximately by 60~MeV towards
the region of smaller $T_p$, as is seen from the comparison of
the curves~1 and~4 in Fig.~6, describing
the $^1S_0$-state contributions to the $pp$-pair formation cross
sections for $E_{pp}=0$ and $E_{pp}=20$~MeV, respectively.
This is a purely kinematical phenomenon connected with increase of
$q'$ (at fixed $T_p$), as $E_{NN}$ is growing.

\section{Relation between the cross sections of the processes of a triplet
$(pn)_t$-pair formation and elastic backward $pd$-scattering}

The kinematics of the reaction $p+d\to p+(pn)$ we deal with
is very close to that one of the elastic backward $pd$-scattering.
If the final $pn$-pair is formed in the triplet state
(presumably in $^3S_1$ at low relative energies),
then it is possible to get some approximate relations between
the cross sections of the processes $p+d\to p+d$ and $p+d\to p+(pn)_{t}$
at high momentum transfers. One of such relations  was
obtained rather long ago by Anisovich, Dakhno and Makarov (ADM)~\cite{adm}
from consideration of the analytical properties of the corresponding
scattering amplitudes. Recently, Boudard, F\"{a}ldt and Wilkin (BFW)
derived another relation~\cite{bfw}, based on analytical continuation of
$^3S_1$-scattering w.f. to the bound state pole.
Both ADM and BFW relations were formulated for a single-channel
problem (i.~e. as if the interaction did not mix channels with different
values of orbital momenta).
One can show, however, that such a restriction is not necessary,
and the similar result holds for a two-channel system as well,
e.~g. for deuteron (see Appendix and Ref.~\cite{sm}).
By this reason, below we will use the subscript "$t$" in the process
$p+d\to p+(pn)_{t}$
as a short notation of the $^3S_1$--${\mathstrut^3D_1}$-state
of the final $pn$-pair.

The ADM and BFW
formulas outwardly look quite different, so, it is interesting
to compare the quality of approximation, which can be reached by
using them. The importance of these relations may become still greater,
if the contribution of the $^3S_1$--${^3D_1}$ final state can be
experimentally separated
anyhow from the total contribution of the singlet and
triplet pair formation (see, for instance, Ref.~\cite{cosy2}).
In the latter case one can make
some estimations of the cross section of the process $p+d\to p+(pn)_{t}$
directly from the data on the well-studied
elastic backward $pd$-scattering.

In an explicit form the ADM and BFW formulas can be written as
\begin{equation}
\label{admbfw}
\frac{d\sigma}{dk^2d\Omega_f}\,(p+d\to p+(pn)_{t})=
\frac{p'_3}{p_1}\,f(k)\,\frac{d\sigma}{d\Omega_f}\,(p+d\to p+d),
\end{equation}
where $d\Omega_f$ is the solid angle element of the backward going proton
in the CMF of the system $p+d$.
According to Ref.~\cite{adm}, the function $f(k)$ is
\begin{equation}
\label{adm}
f_{ADM}(k)={r_t\sqrt{1-{2\, r_t\over a_t}}}
{\left [  2\pi \left (1-
\sqrt{1-{2\, r_t\over a_t}}\right ) \right ]}^{-1}
 {k}{ \left [ k^2+
\left ( -{1\over a_t}+{1\over 2}r_t\, k^2\right )^2\right ]}^{-1}.
\end{equation}
In Eq.~(\ref{adm}) $a_t=5.41$~fm and $r_t=1.75$~fm are the triplet
scattering length and triplet effective radius, respectively.
The BFW prescription
for $f(k)$ looks much simpler:
\begin{equation}
\label{bfw}
f_{BFW}(k)= {k\over{2\pi( k^2+ \alpha_t^2)\alpha_t}}
\end{equation}
with $\alpha_t=\sqrt{m|\varepsilon|}=0.232\,\,\mbox{fm}^{-1}$.
It is very important to emphasize that the relations adduced above
practically do not depend on reaction mechanism, provided the momentum
transfer to the final nucleon pair and to the final deuteron is large enough,
but $k$ is, on the contrary, small.
However, at $k^2>0$ both ADM and BFW relations are approximate.
The reason of it consists in that the exact connection
between the bound state (deuteron) and $pn$-scattering w.f's holds only at
$k^2=k^2_b=-m|\varepsilon|$, separated from the real relative momentum
of the $pn$-pair $k^2=mE_{pn}$ by a finite interval.
Our aim here is to check the validity
of Eq.~(\ref{admbfw}) in concrete kinematical conditions for the case
of the ONE-mechanism.

In order to bring the deuteron break-up cross section into the form
of the left-hand-side of Eq.~(\ref{admbfw}), we must, first of all,
go over from $d^3p'_1$ in Eq.~(\ref{breakupcs}) to $d^3k$,
the relativistic relative momentum ${\bf k}$ being defined by Eq.~(\ref{p}).
The phase space volume element is
\begin{equation}
\label{phasespacevol}
d\Phi=\frac{d^3p'_3}{2E'_3(2\pi)^3}\,\left(\frac{d^3p'_1}{2E'_1(2\pi)^3}
\frac{d^3p'_2}{2E'_2(2\pi)^3}\,\delta^{(4)}(P_i-P_f)\right)\equiv
\frac{d^3p'_3}{2E'_3(2\pi)^3}\,d\Phi_2.
\end{equation}
The two-body phase space volume element $d\Phi_2$ is a relativistic invariant
and can be calculated in any convenient reference frame. We will
choose for finding $d\Phi_2$ the system, where ${\bf p}_1+{\bf P}_d
-{\bf p}'_3=0$. Evidently, this system coincides with the CMF of
the final nucleon pair, where ${\bf p}'_1=-{\bf p'}_2={\bf k}$.
Removing the $\delta$-function, we get
\begin{equation}
\label{phi2}
d\Phi_2=\frac{1}{(2\pi)^6}\,\frac{k\,d\Omega_{\bf k}}{8E_k},
\end{equation}
where $E_k=\sqrt{{\bf k}^2+m^2\mathstrut}$.

Now we find the connection between $p'_3$ and $k$. On the one hand,
\begin{equation}
\label{p3p1}
k^2=\frac{1}{4}\,\left[(E'_1+E'_2)^2-({\bf p}'_1+{\bf p}'_2)^2\right]-m^2.
\end{equation}
Using conservation laws
for the energy and 3-momentum, we can rewrite Eq.~(\ref{p3p1}) in the CMF
of the system $p+d$ as
follows:
\begin{equation}
\label{p3p2}
k^2=\frac{1}{4}\,\left[{s}-3m^2-2\sqrt{s}E'_3\right].
\end{equation}
Here $\sqrt{s}$ is the invariant mass of the system $p+d$.
>From Eq.~(\ref{p3p2}) we easily obtain
\begin{equation}
\label{diffp2}
dk^2={\sqrt{s}\over 2}\,\frac{p'_3dp'_3}{E'_3}.
\end{equation}
Using Eqs.~(\ref{breakupcs}),~(\ref{phasespacevol}),~(\ref{phi2})
and (\ref{diffp2}), we arrive at the following formula for the
cross section\footnote{In Ref.~\cite{imuz90} the phase space volume
was slightly underestimated by the additional factor
{$m/(\sqrt{s}-E_3')$} appeared because of using the nonrelativistic
expression for the relative momentum in the corresponding formula.}:
\begin{equation}
\label{csdp2}
\frac{d\sigma}{dk^2d\Omega_f}={p'_3\over p_1}\,
\frac{k}{(4\pi)^5 \,{s}\, E_k}\,
\int d\Omega_{\bf k}\,\overline{|A_{fi}|^2}.
\end{equation}
We have denoted
$d\Omega'_3=d\Omega_f$ for matching with the corresponding solid angle
element for the case of elastic $pd$-scattering. For identical particles
in the forward going $NN$-pair one has to multiply the right-hand-side of
Eq.~(\ref{csdp2}) by the factor $\frac{1}{2}$.
Now, substituting
the expression~(\ref{amplsquared}) for $\overline{|A_{fi}|^2}$ into
Eq.~(\ref{csdp2}) and applying the partial wave decomposition~(18), we can
easily perform integration over $d\Omega_{\bf k}$ due to the orthogonality
of Legendre polynomials. The result is
$$\frac{d\sigma}{dk^2d\Omega_f}(p+d\to N+(NN))=
{p'_3\over p_1}\,
\frac{k}{(4\pi)^5\, {s}\,E_k}\, {\cal K}\,
\left[u_0^2(q)+u_2^2(q)\right]$$
\begin{equation}
\label{csdp2fin}
\times \sum_{JSLL'}{\cal N}_{NN}(L,S)\,(2J+1) \,|t^{JS}_{LL'}(q',k)|^2,
\end{equation}
where the kinematical factor ${\cal K}$ has the form
\begin{equation}
\label{kin1}
{\cal K}= \frac{E_d\, (E_2+E_3')\,\varepsilon_p(q)}{4\, E_2^2}.
\end{equation}
Note that the formula~(\ref{csdp2fin}) can be derived not only
by using the amplitude of the reaction~(\ref{smu1}),
but also directly from the amplitude of the process
$p+d\to N+(NN)_{JM_JLS}$,
where the state of the final $NN$-pair
is determined by the quantum numbers $J,\,M_J,\,L,\,S$
and by the module of the relative momentum  $k$.

It is worth mentioning that for the reaction
$p+d\to p(180^{\circ})+(pn)$, when only the momentum of the backward
going proton is registered, the kinematics is not collinear and admits
any angle between the vectors ${\bf q}'$ and ${\bf k}$.
At the same time, for the reaction
$p+d\to p(0^{\circ})+n+p(180^{\circ})$ discussed in the previous section
the vector ${\bf k}$ can be directed either along, or oppositely
to the momentum ${\bf q}'$. Therefore, the role of the highest
partial waves in the cross sections of these reactions may be
different.

Since Eq.~(\ref{admbfw}), strictly speaking, is valid for
the triplet $^3S_1$--${^3D_1}$-state of the nucleon pair, we should
retain in the sum over $J$, $S$, $L$, $L'$ only the terms
\eject
\begin{figure}[hbt]
\mbox{\epsfig{figure=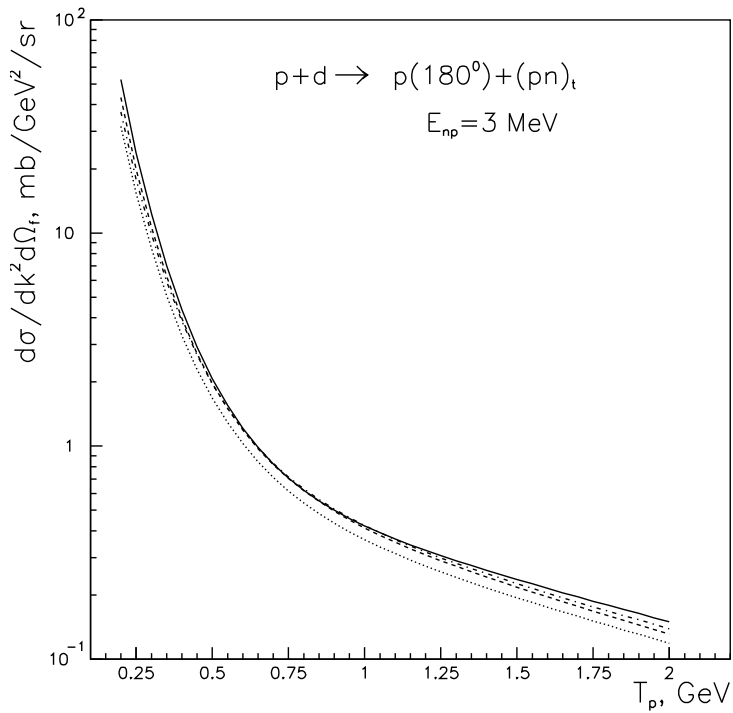,height=0.49\textheight, clip=}}
\caption{ The
CMF differential cross section of the
quasi two-body reaction
$p+d\to p(180^{\circ})+(pn)$ as a function of the initial proton
kinetic energy in the laboratory frame.
The solid line represents the cross section calculated according to
Eq.~(\protect\ref{csdp2fin}) taking into account all the important
states of the final $pn$-pair.
The dashed line describes the cross section obtained
by Eq.~(\protect\ref{csdp2t}) for the triplet
final state $^3S_1$--${^3D_1}$ only.
The dotted and dash-dotted
lines show the right-hand-side of Eq.~(\protect\ref{admbfw}), obtained
by using the functions $f_{ADM}(k)$ (Eq.~(\protect\ref{adm}))
and $f_{BFW}(k)$ (Eq.~(\protect\ref{bfw})), respectively.
All the curves correspond to the relative energy of nucleons
in the $pn$-pair $E_{pn}=3$~MeV.}
\end{figure}
 corresponding
to $J=1$, $S=1$, $L=0,2$ and $L'=0,2$:
$$\frac{d\sigma}{dk^2d\Omega_f}\,(p+d\to p+(pn)_{t})={p'_3\over p_1}\,
\frac{3k}
{(4\pi)^5\,{s}\,E_k}\,{\cal K}\,\left[u_0^2(q)+u_2^2(q)\right]$$
\begin{equation}
\label{csdp2t}
\times\left[|t^{11}_{00}(q',k)|^2+|t^{11}_{20}(q',k)|^2
+|t^{11}_{02}(q',k)|^2+|t^{11}_{22}(q',k)|^2\right].
\end{equation}

The CMF cross section of the elastic $pd$-scattering for the ONE mechanism
is well known in the
relativistic quantum mechanics~\cite{lak} and can be written as
\begin{equation}
\label{sigpddp}
\frac{d\sigma}{d\Omega_f}(p+d\to p+d)={3\over 64\pi^4{s}}\, \Pi^2
\,\left[u_0^2(q)+u_2^2(q)\right]
\left[u_0^2(q')+u_2^2(q')\right],
\end{equation}
where
\begin{equation}
\label{kin2}
\Pi= E_d\,(E_p+E_n)\,\varepsilon_p(q)\,{\sqrt{s}-M_0\over 4\, E_n},
\end{equation}
$E_d=E_d'$ is the energy of the deuteron,
$E_p=E_p'$ is the energy of the proton, $E_n$ is the energy of
the intermediate neutron, $M_0=2E_p+E_n$. The relativistic relative momenta
${\bf q}$ and ${\bf q}'$ are determined by the same
formulas~(\ref{qd}) and~(\ref{k}), respectively,
keeping in mind the difference between the invariant mass of the
 $N_1'N_2'$-pair and the deuteron.
 Note that the equality $q=q'$
takes place in the CMF of the system $p+d$.
\begin{figure}[htb]
\mbox{\epsfig{figure=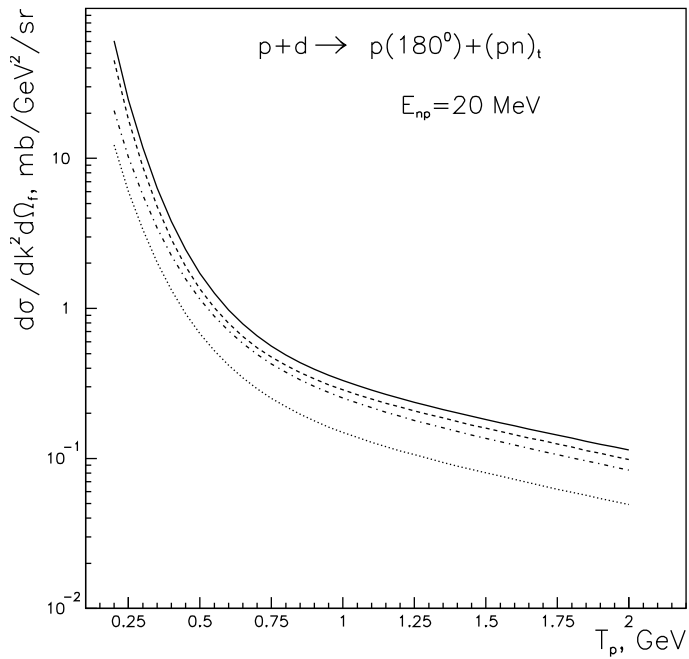,height=0.46\textheight, clip=}}
\caption{ The same as in Fig.~7, but for $E_{pn}=20$~MeV}
\end{figure}

The calculated numerically differential cross sections
$d\sigma/(dk^2\,d\Omega_f)$ of the reaction $p+d\to p(180^{\circ})+(pn)$
as functions of the initial proton kinetic energy $T_p$
are shown in Figs.~7 and~8 for $E_{pn}=3$~MeV and $E_{pn}=20$~MeV,
respectively.
The solid line corresponds to the cross section found by the
formula~(\ref{csdp2fin}) including all the important
partial waves (both singlet and triplet) in the state of the final $pn$-pair.
The dashed line describes the cross section
calculated by Eq.~(\ref{csdp2t}), for only the $^3S_1$--${^3D_1}$-state of the
$pn$-pair taken into account.
For comparison we showed in the same figures the calculated
right-hand-side of Eq.~(\ref{admbfw}), obtained by means of
Eqs.~(\ref{adm}), (\ref{sigpddp}),~(\ref{kin2}) (the dotted line) and
Eqs.~(\ref{bfw}), (\ref{sigpddp}),~(\ref{kin2}) (the dash-dotted line).
>From the figures presented it is seen that Eq.~(\ref{admbfw})
reproduces fairly well the shape of the cross section in the whole
range of initial energies $T_p=0.2$--2~GeV, regardless to the
choice of the function $f(k)$ (ADM or BFW) and to the $pn$-pair excitation
energy $E_{pn}$ in the interval 0--20~MeV.
However, the magnitude of the cross section of $(pn)_t$-pair
formation (the dashed line) can be obtained from
Eq.~(\ref{admbfw}) with an acceptable precision only at rather small
excitation energies. Note that the BFW relation holds with better accuracy
than the ADM one: at $E_{pn}=3$~MeV (Fig.~7)
and $T_p=0.5$--2~GeV the relative error of the former
is about 5\%, while the latter systematically underestimates the cross section
by 10--20\%. The accuracy of Eq.~(\ref{admbfw})
worsens at lower values of $T_p$, because in this region one of the
applicability conditions --- large value of the momentum transfer ---
violates.
When the relative energy $E_{pn}$ increases, the departure from the bound
state (deuteron) pole also increases, and the
accuracy of Eq.~(\ref{admbfw}) falls off.
For instance, at $E_{pn}=20$~MeV (Fig.~8) and
in the same region $T_p=0.5$--2~GeV the BFW relation underestimates the
cross section of a triplet $pn$-pair formation by $\sim 15$\%, whereas the
ADM one --- in about two times.

In paper~\cite{bfw} on the basis of the
relations~(\ref{admbfw}),~(\ref{bfw}) (the BFW recipe)
an attempt was undertaken to separate
contributions of the singlet ($^1S_0$) and triplet {($^3S_1$--${^3D_1}$)}
$pn$-pair states
to the cross section of the deuteron break-up reaction discussed.
But, for experimental resolution of the relative energy
$\Delta E_{pn}\sim 15$--20~MeV, the accuracy of the relation~(\ref{admbfw})
with the function $f_{\rm BFW}(k)$ is about 15\%. The contribution
of the rest partial waves (besides $^1S_0$ and $^3S_1$--${^3D_1}$ ones)
to the cross section gives around 10\% of the latter. Hence, the total
error of separation of the pure $^1S_0$-state contribution by such a
method can be estimated in 25\% of the full cross section.
Therefore, the attempt of the authors of paper~\cite{bfw}
to realize this separation, when the $^1S_0$-state contribution is
expected to be less than 30\% of the full cross section, and at
poor relative energy resolution is, most probably,
an exceeding of the accuracy level of the relation~(\ref{admbfw}).

Nevertheless, it is worth to
mention that at low energy $E_{pn}$ the cross section of
a $^3S_1$--${^3D_1}$-state pair formation accounts for a great deal
of the full cross section given by Eq.~(\ref{csdp2fin}) including
all the partial waves in the $pn$-pair state (cf. the solid and
dashed lines in Figs.~7,~8). Thus, at $E_{pn}=3$~MeV,
the relative weight of the $(pn)_t$-pair contribution in the
full cross section defined by Eq.~(\ref{csdp2fin}) exceeds 90\%.
At $E_{pn}=20$~MeV due to the growing influence of the
highest partial waves this ratio becomes somewhat smaller,
but remains rather high, 75--85\%. So, the ADM and BFW relations can
be used for some estimation of total contribution of the triplet and singlet
pairs to the cross section of deuteron break-up reaction
in quasi-two-body kinematics
\footnote{This statement violates, if $E_{pn}$
is very close to zero, since at such conditions the singlet $^1S_0$-state
dominates.}.

\section*{Conclusion}

We investigated the influence of the off-energy-shell effects
in the $NN$-scattering amplitude on the cross section
of the reaction $p+d \to N(180^o)+(NN)(0^o)$ in the framework
of the ONE mechanism.
For the case of the forward going $pp$-pair a deep minimum in the
corresponding cross section at the initial proton kinetic energy
$T_p\sim 0.77$~GeV is expected, caused by the node of the
${^1S_0}\to {^1S_0}$ transition half-off-shell $NN$-amplitude.
Appearance of such a node is tightly connected with the properties
of the $NN$-interaction in the region of nucleon overlap.
As the relative energy $E_{pp}$ of the protons in the final $pp$-pair
increases, the minimum rapidly fills up because of amplification
of contributions from the highest partial waves.
The position of this minimum was found to be sensitive to
the value of $E_{pp}$.
It was shown that the necessary (but, probably, still not sufficient)
condition of experimental observation of the minimum in question
was high enough relative energy resolution $\Delta E_{pp}\leq 5$~MeV.
Another important feature of the reaction
$p+d \to n(180^{\circ})+(pp)(0^{\circ})$ considered
in the framework of the ONE mechanism
is an approximate constancy of the differential cross section
in the region of the initial energies $T_p\sim 1.4$--2~GeV.

For the reaction of $pn$-pair formation
in the same kinematics of the elastic backward $pd$-scattering
the node of the ${^1S_0}\to {^1S_0}$ transition amplitude
(as well as of the ${^3S_1}\to{^3S_1}$ one)
is not distinctly seen in the corresponding cross section
at any excitation energy
$E_{pn}$, that is caused by significant contribution
of the "nodeless" transition ${^3D_1}\to{^3S_1}$.
However, the nodes discussed may reveal themselves
in polarization phenomena which have been studied so far
only in the impulse approximation (i.~e. without taking
into account the off-shell effects)~\cite{ladygin}.

It was shown also that for the relative energy $E_{pn}\leq 3$~MeV
the cross section of the reaction
$p+d\to p(180^0)+(pn)_t$ of the triplet $pn$-pair formation
can be estimated with
an accuracy $\sim 5$\% by means of the
relations~(\ref{admbfw}),~(\ref{bfw}) based on an
analogy between the processes of
deuteron break-up and elastic backward $pd$-scattering.
Since the latter has been already well studied both theoretically and
experimentally, separation of the channel of the reaction
$p+d\to N+(NN)$ with the singlet $NN$-pair seems mostly interesting.

The authors are grateful to V.~I.~Komarov for his interest to the paper
and stimulating discussions. This work was supported in part
by the Russian Foundation for Basic Research (grant $N^o$ 96-02-17458).

\section*{Appendix}

Below we outline the main steps of proving that Eq.~(\ref{admbfw})
remains approximately valid for deuteron considered as a two-channel
(i.~e. admitting the mixture of $^3S_1$- and $^3D_1$-states) system.
The detailed proof can be found in Ref.~\cite{sm}.

It is more convenient for us to work in the coordinate representation.
The two radial components of the deuteron w.f. in the coordinate space,
$u_0(r)$ and
$u_2(r)$, satisfy the system of two coupled differential equations
(obtained from the Schr\"{o}dinger equation for the initial w.f.):
$$
u''_L(r)-\frac{L(L+1)}{r^2}\,u_L(r)+m\sum_{L''}v_{LL''}(r)u_{L''}(r)-
\alpha_t^2u_L(r)=0\eqno{(A.1)}
$$
and the boundary conditions
$$
u_L(0)=0;\,\,\,\,\,\,u_L(r\to\infty)=N_Le^{-\alpha_t r}.\eqno{(A.2)}
$$
Here $v_{LL''}(r)$ is the radial component of the interaction potential,
$\alpha_t=\sqrt{m|\varepsilon|}$, $N_L$ is some constant.

Let us introduce four radial scattering w.f's of $pn$-system,
${\rm g}_{LL'}(r,k)$, describing transitions between the states
with the "deuteron" quantum numbers, i.~e. $J=1$, $S=1$,
$L=0,\,2$, $L'=0,\,2$. The corresponding system of equations and
boundary conditions for ${\rm g}_{LL'}(r,k)$ are
$$
{\rm g}''_{LL'}(r,k)-\frac{L(L+1)}{r^2}\,{\rm g}_{LL'}(r,k)+
m\sum_{L''}v_{LL''}(r){\rm g}_{L''L'}(r,k)+k^2{\rm g}_{LL'}(r,k)=0,
\eqno{(A.3)}
$$
$$
{\rm g}_{LL'}(0,k)=0;\,\,\,\,\,\,{\rm g}_{LL'}(r\to\infty,k)=
\frac{1}{2ik}\,\left[e^{ikr}\delta_{LL'}-(-1)^{L'}S^*_{LL'}(k)e^{-ikr}\right],
\eqno{(A.4)}
$$
where $S^*_{LL'}(k)$ is the complex conjugated scattering matrix
(for definiteness we consider ${\rm g}_{LL'}(r,k)$ as the
components of the scattering function $\phi^{(-)}({\bf r},{\bf k})$).
The normalization is chosen to be
$$
\sum_L\int^{\infty}_0dr\,\left[u_L(r)\right]^2=1,
\eqno{(A.5)}
$$
$$
\sum_{L}\int^{\infty}_0dr\,{\rm g}_{LL'}^*(r,k'){\rm g}_{LL''}(r,k)
=\frac{\pi}{2k^2}\,\delta_{L'L''}\delta(k-k').
\eqno{(A.6)}
$$

Eqs.~(A.1) and~(A.3) differ from each other
only by the last term on the left-hand-sides.
If we substitute $k$ by $-i\alpha_t$, then the difference disappears,
and the equations become the same. The second boundary condition in Eq.~(A.4)
goes over into
$$
{\rm g}_{LL'}(r\to\infty,k\to -i\alpha_t)=\frac{1}{2\alpha_t}\,
\left[e^{\alpha_t r}\delta_{LL'}-(-1)^{L'}S^*_{LL'}(k\to -i\alpha_t)
e^{-\alpha_t r}\right].\eqno{(A.7)}
$$
Since the $S^*$-matrix
has a pole at $k=-i\alpha_t$, the term with $e^{\alpha_t r}$
is "suppressed", and we get
$$
{\rm g}_{LL'}(r\to \infty,k\to -i\alpha_t) =
\frac{(-1)^{L'+1}}{2\alpha_t}\,S^*_{LL'}(k\to -i\alpha_t)e^{-\alpha_t r},
\eqno{(A.8)}
$$
that up to a constant factor
coincides with the second condition in Eq.~(A.2).
So far as the boundary condition at $r$=0 does not depend on $k$ at all, we
conclude that
$$
u_{L}(r)=\lim_{k\to -i\alpha}\left[A_{L'}(k){\rm g}_{LL'}(r,k)\right],
\eqno{(A.9)}
$$
the coefficient $A$ being independent of $r$ and $L$. In an expanded form
Eq.~(A.9) is equivalent to the two systems of equalities
$$
{u_0(r)\choose u_2(r)}=\lim_{k\to -i\alpha_t}\left[
A_0(k){{\rm g}_{00}(r,k)\choose {\rm g}_{20}(r,k)}\right]
\eqno{(A.10)}
$$
and
$$
{u_0(r)\choose u_2(r)}=\lim_{k\to -i\alpha_t}\left[
A_2(k){{\rm g}_{02}(r,k)\choose {\rm g}_{22}(r,k)}\right].
\eqno{(A.11)}
$$
Note that $A_0(k)$ and $A_2(k)$ are not unique, because they always can be
multiplied by an arbitrary function $\varphi(k)$ satisfying the
condition $\varphi(-i\alpha_t)=1$. One possible choice is~\cite{sm}
$$
A_{0,2}(k)=-\sqrt{\frac{2\alpha_t(k^2+\alpha^2_t)}
{\cos 2\epsilon_1}}\,e^{-i\delta_{0,2}},
\eqno{(A.12)}
$$
where $\delta_{0,2}$ are the scattering phase shifts of the $^3S_1$- and
$^3D_1$-states, respectively, $\epsilon_1$ is the mixing parameter.
All the quantities $\delta_{0,2},\,\epsilon_1$ relate to the
so-called "nuclear bar" parametrization~\cite{stapp}.

In Ref.~\cite{sm} it was shown that Eq.~(A.10) with the coefficient $A_0(k)$
from Eq.~(A.12) holds approximately true, even if we remove the
sign of limit and take the right-hand-side of Eq.~(A.10) at small
real values of $k$. Since under these conditions $|e^{i\delta_0}|=1$,
$\cos 2\epsilon_1\approx 1$, we get
$$
{|u_0(r)|\choose |u_2(r)|}\approx \sqrt{2\alpha_t(k^2+\alpha_t^2)}\,
{|{\rm g}_{00}(r,k)|\choose |{\rm g}_{20}(r,k)|}.
\eqno{(A.13)}
$$
Of course, Eq.~(A.13) is valid with good accuracy, if $k$ is close to the
threshold, and $r<a$, where $a$ is the characteristic radius of the
interaction potential. Going over to the momentum representation
by the formulas
$$
u_L(q')=4\pi(-i)^L\int^{\infty}_0dr\,r\,u_L(r)j_L(q'r),
\eqno{(A.14)}
$$
$$
{\rm g}_{LL'}(q',k)=4\pi(-i)^L\int^{\infty}_0
dr\,r\,{\rm g}_{LL'}(r,k)j_L(q'r)
\eqno{(A.15)}
$$
($j_L(q'r)$ is a spherical Bessel function) and taking into account
the connection between the functions ${\rm g}_{LL'}(q',k)$ and
the previously introduced scattering amplitudes $t^{JS}_{LL'}(q',k)$
$$
{\rm g}_{LL'}(q',k)=\frac{t^{11}_{LL'}(q',k)}{4m(q'^2-k^2-i0)},
\eqno{(A.16)}
$$
we obtain
$$
(q'^2-k^2)^2\left[u_0^2(q')+u_2^2(q')\right]\approx
\frac{\alpha_t(k^2+\alpha^2_t)}{8m^2}\,\left[|t^{11}_{00}(q',k)|^2+
|t^{11}_{20}(q',k)|^2\right].
\eqno{(A.17)}
$$
Although, strictly speaking, Eq.~(A.11) can not be treated in the same
way, as Eq.~(A.10), and the relation between $u_0(q')$, $u_2(q')$
and the amplitudes $t^{11}_{02}(q',k)$, $t^{11}_{22}(q',k)$,
analogous to Eq.~(A.17),
does not exist, at $ka\ll 1$
the sum $|t^{11}_{02}(q',k)|^2+|t^{11}_{22}(q',k)|^2$
is small compared to
$|t^{11}_{00}(q',k)|^2+|t^{11}_{20}(q',k)|^2$. Therefore, without
significant change of accuracy we may extend Eq.~(A.17) to
include into its right-hand-side all the four amplitudes:
$$
(q'^2-k^2)^2\left[u_0^2(q')+u_2^2(q')\right]\approx
\frac{\alpha_t(k^2+\alpha^2_t)}{8m^2}
\,\left[|t^{11}_{00}(q',k)|^2+
|t^{11}_{20}(q',k)|^2\right.$$
$$\left.+|t^{11}_{02}(q',k)|^2+|t^{11}_{22}(q',k)|^2
\right].
\eqno{(A.18)}
$$
Numerical estimations show that even at $k^2/m=20$~MeV the
relative difference between the right-hand-sides of
Eqs.~(A.17) and~(A.18) is less than 6\%, when $q'$
varies from 0 up to 1~GeV.

Now, substituting Eq.~(A.18) into Eq.~(\ref{sigpddp}) and comparing
the result with Eq.~(\ref{csdp2t}) (keeping in mind that
$q'^2\gg k^2\sim m|\varepsilon|$), we reproduce in the nonrelativistic limit
Eq.~(\ref{admbfw}) with
the function $f_{BFW}(k)$ defined by Eq.~(\ref{bfw}).

On the basis of the same arguments we conclude that Eq.~(\ref{admbfw})
with the function $f_{ADM}(k)$ (Eq.~(\ref{adm})), being
true for the single $^3S_1$-state, holds also
for the case of coupled $^3S_1$--${^3D_1}$ channels.

}

\end{document}